\documentclass{iau307}
\usepackage{graphicx}
\usepackage{natbib}
\usepackage{url}
\usepackage{dtklogos}
\bibpunct{(}{)}{;}{a}{}{,}

\title[Cepheid RV Modulation] 
{Amplitude Modulation of Cepheid Radial Velocity Curves as a Systematic Source
of Uncertainty for Baade-Wesselink Distances}

\author[R.I.~Anderson]   
{Richard I. Anderson$^1$
}

\affiliation{$^1$Observatoire de Gen\`eve, Universit\'e de Gen\`eve, 51 Ch. des
   Maillettes, 1290 Sauverny, Switzerland \\ [\affilskip] email: {\tt
   richard.anderson@unige.ch}}

\pubyear{2014}
\volume{307} 
\pagerange{}
\jname{New windows on massive stars: asteroseismology, interferometry, and spectropolarimetry}
\editors{G. Meynet, C. Georgy, J.H. Groh \& Ph. Stee, eds.}

\begin{document}

\maketitle

\begin{abstract}
I report on the recent discovery of modulation in the radial velocity curves
in four classical Cepheids. This discovery may enable significant improvements
in the accuracy of Baade-Wesselink distances by revealing a not previously
considered systematic source of uncertainty.  


\end{abstract}

\firstsection 
\section{Introduction}
Cepheids are excellent standard candles thanks to a relation between the
logarithm of the period and their intrinsic brightness
\citep{1912HarCi.173....1L}.
This period-luminosity relation (PLR) finds application on Galactic and
extragalactic scales and is crucial to precisely measure the expansion rate
of the universe \citep{2011ApJ...730..119R}. 

Baade-Wesselink (BW) type methods determine geometric distances by comparing
angular and linear radius variations due to pulsation.
They have been applied both to Cepheids in the Galaxy and the Magellanic Clouds
\citep[see][and references therein]{2013IAUS..289..138G}. Hence, different
metallicities are probed using a homogeneous method, which is of great
value regarding a possible metallicity dependence of the PLR. However, the accuracy of BW-type
distances has suffered from certain difficulties usually ascribed to the
projection (p-)factor \citep[e.g.][]{2009A&A...502..951N}, which is used to
translate measured radial velocity into pulsation velocity.

\section{Radial Velocity Amplitude Modulations}
\begin{figure}
\centering
\includegraphics[width=\textwidth]{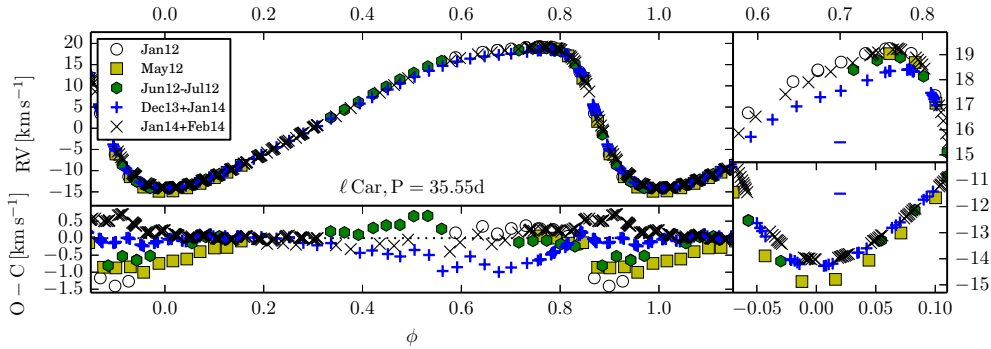}
\caption{The modulated radial velocity curve of $\ell$~Carinae. Figure from
\cite{2014A&A...566L..10A}.}
\label{fig:lCar}
\end{figure}

Using high-precision radial velocities (RVs) obtained with the {\it Coralie}
spectrograph mounted to the Swiss Euler telescope at La Silla,
\citet{2014A&A...566L..10A} was recently able to show that Cepheid radial
velocities are subject to modulation. This was demonstrated for two
short-period, as well as two long-period Cepheids. Figure\,\ref{fig:lCar}
exemplifies this phenomenon for the long-period $\ell$\,Carinae. 

\section{A Systematic Uncertainty for BW distances}
BW-type methods determine distance from the ratio of linear ($\Delta R$) and
angular ($\Delta \theta$) radius variations due to pulsation. $\Delta R$ is
computed as the integral of the pulsational velocity curve and therefore depends
on the epoch of observation, if modulation is present. In consequence, combining
measurements of $\Delta \theta$ from one epoch with $\Delta R$ from another
epoch can lead to systematic errors in the derived distance, reaching up to
$5-15\%$ \citep{2014A&A...566L..10A}.
It is thus crucial to measure angular and linear radius variations
contemporaneously (equivalent pulsation cycles).
While the impact of modulation on the distance estimate may average out over
very long timescales (if the mean RV curve is stable), it appears likely that
modulation explains a good portion of what has thus far limited the accuracy of
BW distances.

\section{The way forward}
There are several possible explanations for the origin of RV modulations. From
the fact that modulations in short-period Cepheids appear steady over longer
timescales (years), while long-period Cepheids exhibit seemingly stochastic
(cycle-to-cycle) modulation, it appears likely that different effects are
responsible for the observed phenomenon. A longer observational baseline is
needed to investigate possible periodicity and the frequency of the phenomenon.

Observed period jitter in the {\it Kepler} Cepheid V1154\,Cygni
\citep{2012MNRAS.425.1312D} and {\it flickering} observed in two Cepheids with {\it MOST}
\citep{2014IAUS..301...55E} suggests that RV modulations have a photometric
counterpart.
However, no Cepheid has as of yet been found to exhibit both RV modulation and
photometric flickering. Only the long-period Cepheid RS\,Puppis exhibits both
strong stochastic period variations and RV modulation. Both phenomena  
are obvious in the unprecedented {\it Coralie} RV data set.

In conclusion, significant improvements in BW distance accuracy may be achieved
by determining linear and angular radius variations contemporaneously, if
RV curve modulations directly correlate with modulations in angular diameter.
Dense contemporaneous time-series of high-quality spectroscopic, photometric,
and interferometric measurements are required to investigate this exciting
possibility.

\bibliographystyle{iau307}
\bibliography{MyBiblio}

\end{document}